\documentclass{article}

\usepackage{lipsum} 

\usepackage[sc]{mathpazo} 
\usepackage[T1]{fontenc} 
\usepackage{microtype} 
\usepackage[hmarginratio=1:1,top=25.4mm,bottom=25.4mm,left=25.4mm,right=25.4mm,columnsep=20pt]{geometry}
\usepackage{multicol} 
\usepackage[hang, small,labelfont=bf,up,textfont=it,up]{caption} 
\usepackage{booktabs} 
\usepackage{hyperref} 

\usepackage{lettrine} 
\usepackage{paralist} 

\usepackage{abstract} 

\usepackage{fancyhdr} 
\usepackage{amsmath}
\usepackage{graphicx}
\usepackage{caption}
\usepackage{subcaption}
\usepackage{amsfonts}
\usepackage{rotating}
\usepackage{amsthm}
\usepackage{MnSymbol}
\usepackage{tikz}

\usetikzlibrary{arrows}
\usepackage{wasysym}

\usepackage{comment}

\title{
A dynamical systems model of unorganized segregation}

\author{D. J. Haw\thanks{D. J. Haw, School of Public Health, Imperial College London. S. J. Hogan: Department of Engineering Mathematics, University of Bristol, Bristol BS8 1UB} and S. J. Hogan}

\date{}
\begin{document}

\maketitle

\begin{abstract}
We consider Schelling's bounded neighbourhood model (BNM) of unorganised segregation of two populations from the perspective of modern dynamical systems theory. We derive a Schelling dynamical system and carry out a complete quantitative analysis of the system for the case of a linear tolerance schedule in both populations.

In doing so, we recover and generalise Schelling's qualitative results. For the case of unlimited population movement, we derive exact formulae for regions in parameter space where stable integrated population mixes can occur. We show how neighbourhood tipping can be adequately explained in terms of basins of attraction.

For the case of limiting population movement, we derive exact criteria for the occurrence of new population mixes and identify the stable cases. We show how to apply our methodology to nonlinear tolerance schedules, illustrating our approach with numerical simulations.

We associate each term in our Schelling dynamical system with a social meaning. In particular we show that the dynamics of one population in the presence of another can be summarised as follows
\begin{multline*}
\{\mbox{rate of population change}\} = \{\mbox{intrinsic popularity of neighbourhood}\} - \{\mbox{finite size of neighbourhood}\} \\
-\{\mbox{presence of other population}\}
\end{multline*}
By approaching the dynamics from this perspective, we have a complementary approach to that of the tolerance schedule.
\end{abstract}


\section{Models of segregation}
Segregation is "the action or state of setting someone or something apart from others or the enforced separation of different racial groups in a country, community, or establishment" (collectively known as {\it organized} segregation) or it can occur as the result of the "interplay of individual choices" (known as {\it unorganized} segregation) \cite{Schelling1969, Schelling1971}. In terms of societal, political and economic outcomes, segregation is widely regarded as undesirable. It is the opposite of integration.

Schelling's {\it spatial proximity model} (SPM) \cite{Schelling1969, Schelling1971} was the first model of unorganized segregation. It is a discrete-time spatial (agent-based) model that uses a chequerboard framework in which cells (representing physical units such as a house in a street or a bed in a dormitory) are occupied - or not - by equal numbers of two different types of agents. At each time-step, agents remain where they are unless the proportion of agents of the other type in their neighbourhood exceeds a given threshold, in which case they move to a vacant cell. Many variants of this model exist, including different group sizes and tolerance demands \cite{Singh2011a}, different methods of relocation \cite{ Laurie2003, Pollicott2001, Zhang2009}, non-lattice topologies \cite{Henry2011a, Pancs2007}, and simulations based on demographic and geographical data (\cite{Benenson1998, Benenson1999, Benenson2013, Benenson2002, Hatna2012}). The emergent behaviour of all such models is the same: even in highly-tolerant populations, a small preference for familiarity in one's immediate neighbours is sufficient to induce geographical segregation \cite{Fossett2006a, Fossett2005, Henry2011a, Pancs2007}.


In the same papers, Schelling \cite{Schelling1969, Schelling1971} also introduced the {\it bounded neighbourhood model} (BNM), which has been seldom pursued in the literature \cite{Dodson2014, Clark1991}. It is the purpose of this paper to examine Schelling's BNM within the framework of modern dynamical systems theory \cite{StrogatzBook}. In doing so, we recover Schelling's results analytically, generalise them and give some new results.


%
%
\section{Schelling's bounded neighbourhood model (BNM)}
\label{sec:BNM}
In Schelling's bounded neighbourhood model (BNM) \cite{Schelling1969, Schelling1971}, the population is divided into two types. Both populations can have different {\it tolerance limits}. A neighbourhood is like a district within a city, or a workplace/social group. An individual in that neighbourhood moves out if they are not happy with the population mix there. 
In the neighbourhood, every member of the population is concerned about the distribution of the types of agents, not with any particular configuration. 

Let $X(t)\geq 0$ and $Y(t)\geq 0$ denote the density of two population types inhabiting the neighbourhood, as a function of time $t$. Note that Schelling \cite{Schelling1969, Schelling1971} refers to $W$ (Whites) and $B$ (Blacks). 

In the neighbourhood, tolerance limits are allocated to a given population type via a {\it tolerance schedule}. The $X$-type tolerance schedule $R_X(X)$ describes the minimum ratio $X/Y$ required in order for all of the $X$ population to remain in that neighbourhood. A similar function $R_Y(Y)$ describes the tolerance of the $Y$-type population. 
Schelling \cite{Schelling1969, Schelling1971} made the following assumptions:
\begin{enumerate}
\item The neighbourhood is preferred over other locations: populations of both type will enter/remain unless tolerance conditions are violated. 
\item The tolerance schedule is specific to the location being studied. 
\item Each member of the population is aware of the ratio of population types within the neighbourhood at the moment the decision is made to enter/remain/leave. 
\item There is no lower bound on tolerance, that is, no population {\it insists} on the presence of the opposing type. 
\item Tolerance schedules are monotone decreasing, i.e. the more tolerant population is first to enter and last to leave. 
\end{enumerate}

\begin{figure}[ht]
\centering
\begin{subfigure}{.45\textwidth}
\centering
\includegraphics[width=\linewidth]{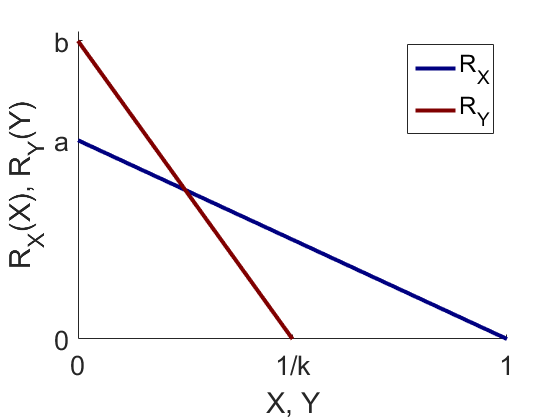}
\caption{Linear tolerance schedules}
\label{fig:tol}
\end{subfigure}\qquad\qquad
\begin{subfigure}{.45\textwidth}
\centering
\includegraphics[width=\linewidth]{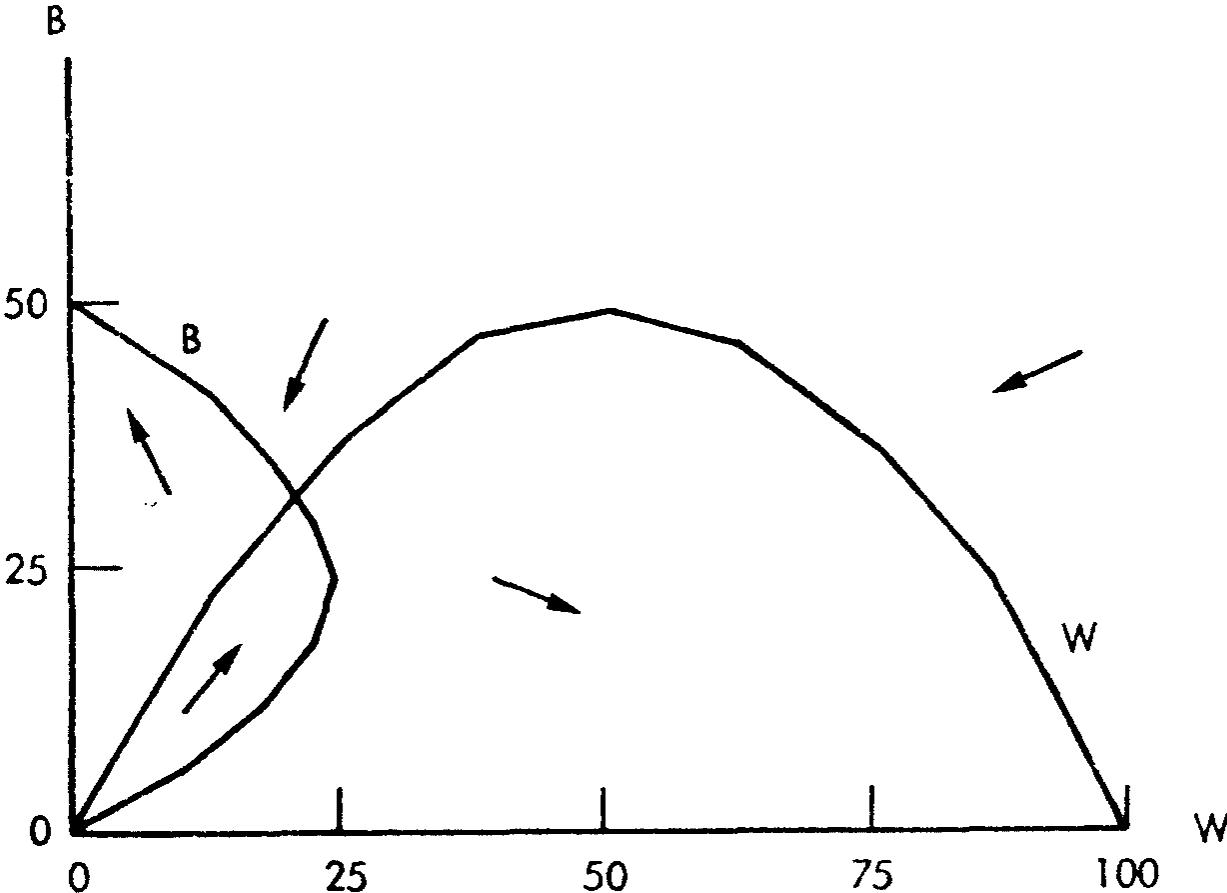}
\caption{A reproduction of \cite{Schelling1971}, Figure $18$.}
\label{SchPP}
\end{subfigure}
\caption{Schelling's first example: linear tolerance schedules and their translation the phase plane.}
\label{fig:fig1}
\end{figure}

Schelling's initial example of a tolerance schedule is linear, as shown in Figure \ref{fig:tol}. In this paper, we take units of population density such that $X_{max}=1=kY_{max}$ for some $k>0$ (Schelling initially sets $X_{max}=100$ and $Y_{max}=50$, so $k=2$). The most tolerant member of the $X$-population can abide a $Y:X$ ratio of $a:1$, half of the $X$-population can tolerate a ratio of $a/2:1$, and the least tolerant member of the $X$-population can abide no members of the $Y$-population. Likewise, the most tolerant member of the $Y$-population can abide a $X:Y$ ratio of $b:1$ etc. We have $a,b>0$, and we can assume that $k\geq 1$, so that $Y$ denotes the minority population.  

Each population has its own tolerance limit. For the $X$-population, this is the value of $Y$-population {\it above which} the $X$-population will leave and {\it below which} there will be an $X$-population influx. 

These limits correspond to $Y/X=R_X(X)$ and $X/Y=R_Y(Y)$. Therefore, for linear tolerance schedules, these limits are parabolas in the $(X,Y)$ plane, given by
\begin{eqnarray}
Y&=&XR_X(X)=a X(1-X) \label{parab1}\\
X&=&YR_Y(Y)=b Y(1-kY) \label{parab2}
\end{eqnarray}
Figure~\ref{SchPP} is Schelling's own sketch (\cite[Figure 18]{Schelling1971}, see also \cite[Figure 1]{Schelling1969}) of these parabolas, together with arrows that give an indication of the qualitative dynamics that arise. In modern dynamical systems nomenclature, Figure~\ref{SchPP} is a plot of the system dynamics in the $(W,B)$ {\it phase plane}. In our notation, Figure~\ref{SchPP} corresponds to $a=b=k=2$.

Schelling explained Figure~\ref{SchPP} in terms of {\it static viability} and {\it dynamics of movement} and identified two {\it stable equilibria}. 
\begin{itemize}
\item Static viability: Any point that lies within the area of overlap of the two parabolas denotes a {\it statically viable} combination of the $W$ and $B$ populations. Any other point between the $W$-parabola  and the $W$-axis represents a mixture of $W$ and $B$ where all the $W$, but not all the $B$, will remain. Similarly any point outside the area of overlap but between the $B$-parabola and the $B$-axis represents a mixture where all the $B$, but not all the $W$, will remain. Any point outside both parabolas denotes a mixture of $W$ and $B$ where neither all of $W$ nor all or $B$ are satisfied.
\item Dynamics of movement: Schelling argued qualitatively that outside the area of overlap, some of $W$ or $B$ are unhappy, so they will move. The area of overlap itself is attractive and so will lead to an influx of people and hence instability.
\item Stable equilibria: Schelling argued that, for the example in Figure~\ref{SchPP}, there are two stable equilibria, both corresponding to a segregated population: the area would contain {\it either} all $W$ and no $B$ {\it or} it would be all $B$ and no $W$ (these are the points of intersection of the parabola with the axes). The statically viable points are not stable, in this case.
\end{itemize}
Schelling gave another example (\cite[Figure 19]{Schelling1971}, see also \cite[Figure 2]{Schelling1969}), for different parameter values, which resulted in an integrated population. Segregation is also possible in this second case, but the eventual population mix depends on the initial values of $X(t),Y(t)$.

Schelling considered the following extensions of the BNM: 
\begin{enumerate}
\item A limit on the number of one of the two types of population, but not both.
\item A limit on the $X$-population {\it and} a limit on the $Y$-population.
\item A limit on the total population.
\item A range of tolerance schedules.
\item Limiting the ratio $X/Y$.
\item The effect of perturbations in the system, such as when a group of one type enters or leaves the area. When such behaviour changes the equilibrium state to which the system converges, this is known as {\it neighbourhood tipping} \cite[p.181]{Schelling1971}.
\end{enumerate}

In this paper, we consider $1,\ 2,\ 4$ and $6$ in detail, owing to their analytic tractability. We give analytic criteria to determine whether equilibria are stable or not and show how limiting a population can introduce new, stable, equilibria. Extensions 3 and 5 can be considered numerically, using the same methods. 

\section{Methods}
Schelling's arguments are qualitative. 
But modern dynamical systems theory \cite{StrogatzBook} provides the ideal framework to develop a fully predictive model, that we call a {\it Schelling dynamical system}. Suppose we have an arbitrary dynamical system in two time-dependent population variables $X(t), Y(t)$ of the form
\begin{align}\label{gendynsys}
\frac{dX}{dt} & \equiv  \dot X  =   F(X,Y)\\
\frac{dY}{dt} & \equiv  \dot Y =   G(X,Y)\nonumber
\end{align}
Then the {\it nullclines} of \eqref{gendynsys} are given by the curves $F(X,Y)=0$ (when there is no growth in the $X$-population) and $G(X,Y)=0$ (when there is no growth in the $Y$-population). Nullclines correspond to curves in $(X,Y)$-phase plane with the same (zero) slope. The {\it intersection} of nullclines gives the {\it equilibria} (or {\it fixed points}) of \eqref{gendynsys}, whose stability can then be examined.

Our observation is that the parabolas in Figure~\ref{SchPP} correspond to the $X$- and $Y$-\textit{nullclines} of a dynamical system within the given neighbourhood. In addition, the lines $X=0$ and $Y=0$ are nullclines. 
Then the Schelling dynamical system for a linear tolerance schedule is given by 
\begin{eqnarray}
\dot X&=&\left[aX(1-X)-Y\right]X \label{DS}\\
\dot Y&=&\left[bY(1-kY)-X\right]Y. \nonumber
\end{eqnarray}
We note that \eqref{DS} is very similar to a Lotka-Volterra system \cite{StrogatzBook}, where growth and decay terms compete to determine the overall population dynamics. Both $a$ and $b$ have units $T^{-1}$. Also \eqref{DS} automatically satisfies Assumption 1 of section \ref{sec:BNM} (see discussion below).

Note that we can rescale time $\hat t = at$ and set $Y=aZ$. Then \eqref{DS} becomes
\begin{eqnarray}
\dot X&=&\left[X(1-X)-Z\right]X \label{nonDS}\\
\dot Z&=&\left[\beta Z(1-\alpha Z)-X\right]Z \nonumber
\end{eqnarray}
where 
\begin{eqnarray}\label{alphabeta}
\alpha \equiv ak > 0, \quad \beta \equiv ab > 0.
\end{eqnarray}

We shall study the dynamics of \eqref{nonDS} for arbitrary values of $\alpha, \beta$, referring to \eqref{DS} and $a,b,k$ whenever necessary. Mathematically, it is more natural to work with the $(X,Z)$ variables. But practically, we are interested in the behaviour of the $(X,Y)$ variables, as was Schelling. So we will alternate between their usage, depending on the context. Our methods are analytical where possible and numerical where necessary.

\section{Unlimited numbers}\label{sec:unlimited}
We  begin with the case referred to by Schelling as {\it unlimited numbers}. Here the neighbourhood can take up to the maximum amount of both populations, so we can have $X_{max}=1$ or $Y_{max}=\frac{1}{k}$, $(Z_{max}=\frac{1}{\alpha})$ within the neighbourhood. We then ask: 
\begin{itemize}
\item What are the equilibria of \eqref{nonDS}?
\item Under what conditions are the equilibria of \eqref{nonDS} stable? 
\item How segregated/integrated are such stable equilibria?
\item For which \textit{initial values} of $X$ and $Z$ does the system converge to a stable equilibrium, for fixed values of $\alpha,\beta$?
\end{itemize}\par 

\subsection{Equilibria}
Equlibria, or steady state solutions, correspond to $\dot{X}=\dot{Z}=0$. So $(X,Z)=(X_e,Z_e) = (0,0), (1,0),(0,\frac{1}{\alpha})$ are equilibria of  \eqref{DS}. These correspond, respectively, to: a) the neighbourhood is empty of both populations, b) the neighbourhood consists of the $X$-population only and c) the neighbourhood consists of the $Z$-population only.

\begin{figure}[h!]
\centering
\includegraphics[width=0.5\linewidth]{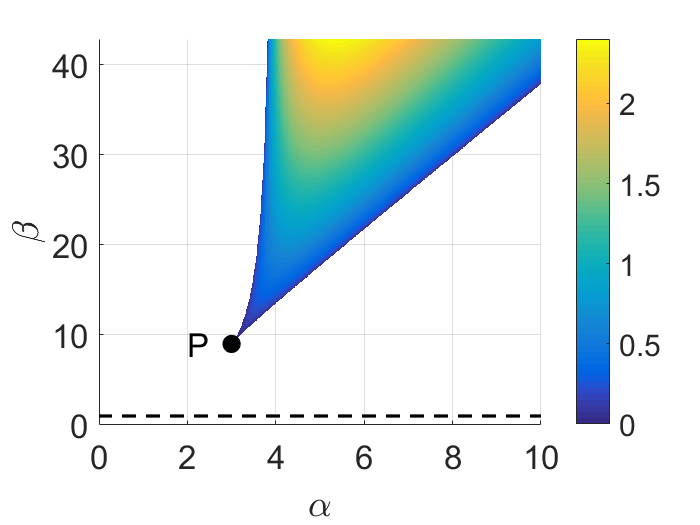}
\caption{Equation \eqref{fixpoint} has three real roots in the coloured area and one real root outside (and above the dotted line $\beta=1$). The upper branch is the curve $\beta = \beta_+$, which has a vertical asymptote at $\alpha = 4$. The lower branch is the curve $\beta = \beta_-$. The apex of the coloured region is the point $P$: $(\alpha,\beta) = (ak, ab) = (3,9)$. Below the dashed line $\beta=1$, equation \eqref{fixpoint} has one real root $X_e < 0$. The colour scale shows $det(J)$ evaluated at the equilibrium point corresponding to the intermediate real root of equation \eqref{fixpoint}, where this root exists.}
\label{fig:alphabeta}
\end{figure}

There is also the possibility of other equilibria, corresponding to the intersection of the nullclines $Z=X(1-X)$ and $X=\beta Z(1-\alpha Z)$, which happens when there is at least one real root $(X,Z)=(X_e,Z_e)$ of the cubic equation
\begin{eqnarray}\label{fixpoint}
X_e^3+a_2X_e^2+a_1X_e+a_0&=&0
\end{eqnarray}
where $Z_e=X_e(1-X_e)$ and we have set 
\begin{eqnarray}\label{an}
a_2 \equiv -2, \quad a_1 \equiv \frac{1+\alpha}{\alpha}, \quad a_0 \equiv \frac{1-\beta}{\alpha\beta}.
\end{eqnarray}
We are interested in real (positive) values of $(X_e, Z_e)$. A cubic equation with real coefficients such as \eqref{fixpoint} has three either one real root and two complex roots or three real roots. In order to distinguish between the two possibilities, we must calculate the {\it discriminant} $D$ of the cubic in \eqref{fixpoint}. Then if $D>0$, \eqref{fixpoint} has one real root and if $D<0$, \eqref{fixpoint} has three real roots. In \eqref{fixpoint}, $D = a_2^2a_1^2-4a_1^3-4a_2^3a_0-27a_0^2+18a_2a_1a_0$, so we have
\begin{eqnarray}\label{D}
D & = & \frac{[(4-\alpha)\beta^2+(4\alpha^2-18\alpha)\beta+27\alpha]}{108\alpha^3\beta^2}
\end{eqnarray} 
and hence \eqref{fixpoint} has three real roots when
\begin{equation}\label{beta}
\beta_-(\alpha) < \beta < \beta_+(\alpha)
\end{equation} 
where
\begin{equation}\label{betapm}
\beta_{\pm}(\alpha) =  \frac{9\alpha-2\alpha^2 \pm 2\sqrt{\alpha(\alpha-3)^3}}{4-\alpha}
\end{equation} 
provided
\begin{equation}\label{alpha}
\alpha  >  3.
\end{equation} 
Thus, we will have three real roots when $(\alpha,\beta) \equiv (ak, ab)$ lies in the shaded region of Figure~\ref{fig:alphabeta}.

In terms of the original parameters $a,b,k$, we will have three real roots for equation \eqref{fixpoint} when 
\begin{equation}\label{b}
b_-(ak) < b < b_+(ak)
\end{equation} 
where
\begin{equation}\label{bpm}
b_{\pm}(ak) =  \frac{9ak-2a^2k^2 \pm 2\sqrt{ak(ak-3)^3}}{a(4-ak)}
\end{equation} 
provided
\begin{equation}\label{ak}
ak  >  3.
\end{equation} 
Note that when $\beta \equiv ab \le 1$, we have only one real root $X_e \le 0$, which is unphysical.

\subsection{Stability analysis}\label{subsec:stab}
The stability of a equilibrium is determined by the eigenvalues $\lambda_1,\lambda_2$ of the Jacobian for the system evaluated at that equilibrium. The Jacobian of \eqref{DS} is given by
\begin{eqnarray}\label{jac}
J(X,Z)=\left( \begin{array}{cc}
X(2-3X)-Z & -X \\
-Z & \beta Z(2-3\alpha Z)-X \end{array} \right)
\end{eqnarray}
For the equilibrium $(X_e, Z_e) = (0,0)$, we find that the eigenvalues of $J(0,0)$ are both zero. So the stability of this equilibrium is determined by the type of perturbation. But since it corresponds to an empty neighbourhood, we focus attention on the remaining equilibria.

We next consider the segregated equilibria $(1,0)$ and $(0,\frac{1}{\alpha})$. 
In both cases, both eigenvalues are negative. For $(X_e, Z_e) = (1,0)$, the (degenerate) eigenvalues are $-1$, $-1$. This equilibrium is a stable (degenerate, or improper) node. The eigenvector is the $X$-axis. Hence the importance of taking $Y=0 \enspace (Z=0)$ as a nullcline. If the initial population consists of members of the $X$-population only, then they will attract more members of the same type until $(X, Z) = (X_e, Z_e) = (1,0)$, that is the neighbourhood is filled with the $X$-population. For the other equilibrium $(X_e, Z_e) = (0,\frac{1}{\alpha})$, the eigenvalues are $-\frac{1}{\alpha}$, $-\frac{\beta}{\alpha}$. This equilibrium is a stable node. It has eigenvectors spanned by $(1,\frac{1}{1-\beta})$ and the line $X=0$. Hence the importance of taking $X=0$ as a nullcline. 

Let us now consider the equilibrium given by the intermediate solution $(X_e,Z_e)$ of the cubic in \eqref{fixpoint}, when $3$ real solutions exist.
Analytically, this case is a lot harder than the others, so we present our results numerically. For each point $(\alpha,\beta)$ in Figure \ref{fig:alphabeta}, we calculate $X_e$. 
Then, since $Z_e=X_e(1-X_e)$, we can evaluate $J(X_e,Z_e)$ using \eqref{jac}. We plot $det(J)$ as a function of $(\alpha,\beta)$ in Figure \ref{fig:alphabeta}. We have $det(J)>0$ in the whole coloured region, i.e. where \eqref{fixpoint} has $3$ distinct real roots. Since $det(J)=\lambda_1\lambda_2$, both eigenvalues are either positive or negative here. Simple inspection shows that both eigenvalues are negative, so the solution is stable. Similarly, we can show that the other $2$ real solutions to~\eqref{fixpoint} have $det(J)<0$ and so are both saddles. Outside the coloured region, we have a single real solution with $det(J)<0$ so the single real solution is a saddle for these values of $(\alpha,\beta)$.


\subsection{Phase portraits and bifurcation diagrams}
\begin{figure}[h!]
\centering
\begin{subfigure}{.45\textwidth}
\centering
\includegraphics[width=\linewidth]{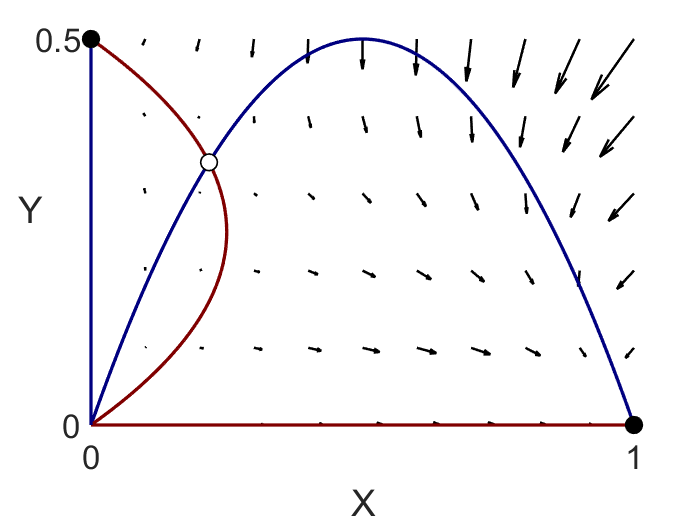}
\caption{$(a,b,k)=(2,2,2)$; $(\alpha,\beta) = (4,4)$}
\label{bif0}
\end{subfigure}\qquad\qquad
\begin{subfigure}{.45\textwidth}
\centering
\includegraphics[width=\linewidth]{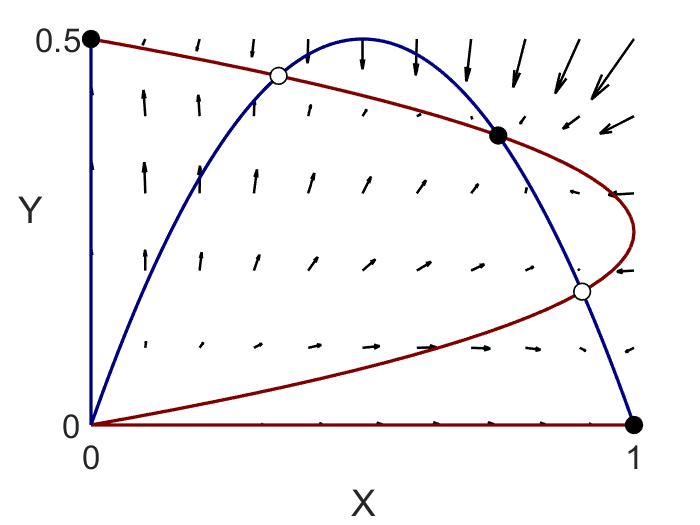}
\caption{$(a,b,k)=(2,8,2)$; $(\alpha,\beta) = (4,16)$}
\label{bif1}
\end{subfigure} \\
\begin{subfigure}{.45\textwidth} 
\centering
\includegraphics[width=\linewidth]{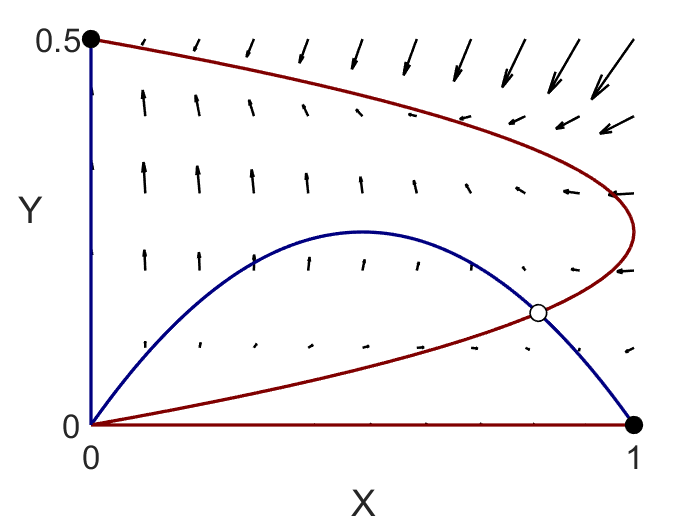}
\caption{$(a,b,k)=(1,8,2)$; $(\alpha,\beta) = (2,8)$}
\label{bif2}
\end{subfigure}\qquad\qquad
\begin{subfigure}{.45\textwidth} 
\centering
\includegraphics[width=\linewidth]{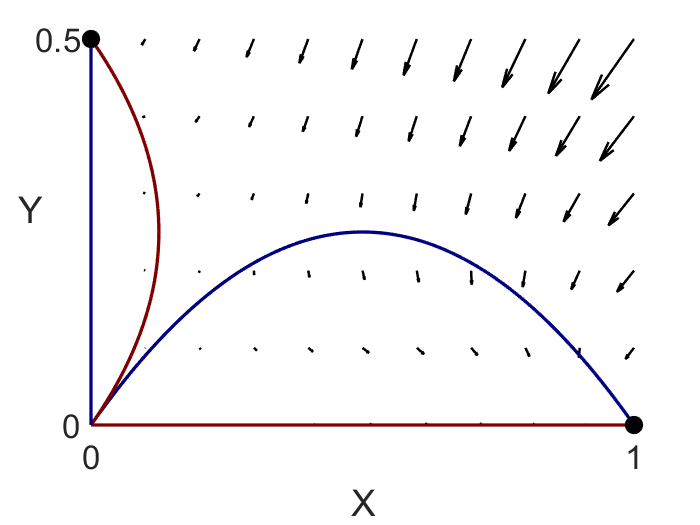}
\caption{$(a,b,k)=(1,1,2)$; $(\alpha,\beta) = (2,1)$}
\label{bif3}
\end{subfigure}
\caption{Phase portraits corresponding to different points in Figure~\ref{fig:alphabeta}. Stable equilibria are shown as filled circles, and saddle points as open circles.}
\label{phaseports}
\end{figure}

Figure \ref{phaseports} shows phase portraits, now in the $(X,Y)$ plane, for different values of $(a,b,k)$, together with the corresponding values of $(\alpha,\beta)$. Setting $(a,b,k)=(2,2,2)$, as in Schelling's initial example (our Figure~\ref{SchPP}), we obtain the phase portrait shown in Figure~\ref{bif0}. We are in the unshaded area of Figure~\ref{fig:alphabeta}.

We have the same qualitative behaviour as in Figure~\ref{SchPP}. Figure~\ref{bif0} also contains quantitative information about the local direction of movement of the population mix. The bigger the arrow, the faster the movement. So we see that, in this case, movement toward a segregated $X$-population is much faster than towards a segregated $Y$-population. Movement in the neighbourhood of the fourth equilibrium (a saddle) is almost imperceptible.

\begin{figure}[h!]
\centering
\begin{subfigure}{.45\textwidth}
\centering
\includegraphics[width=\linewidth]{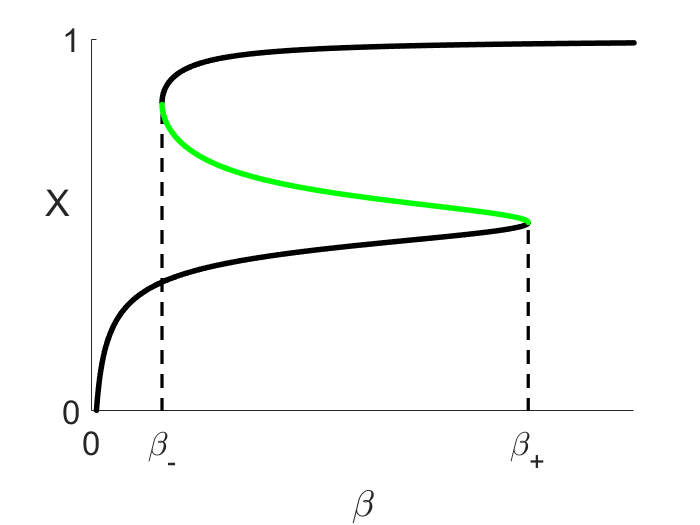}
\caption{$3<\alpha<4$}
\label{bifcontk1}
\end{subfigure}\qquad\qquad
\begin{subfigure}{.45\textwidth}
\centering
\includegraphics[width=\linewidth]{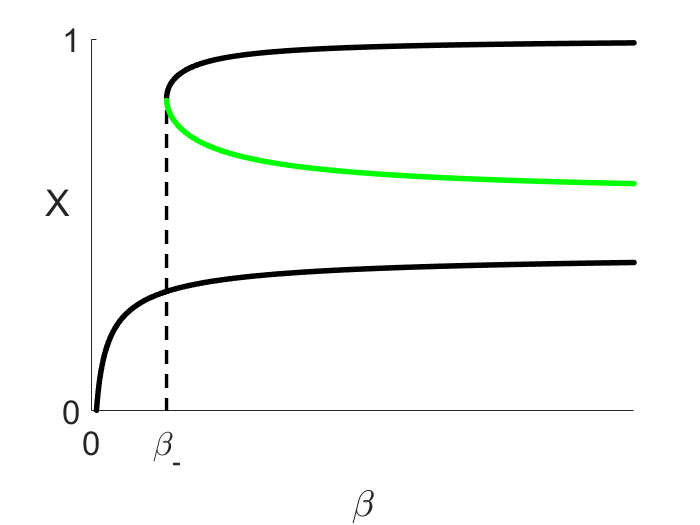}
\caption{$\alpha>4$}
\label{bifcontk1_1}
\end{subfigure}
\caption{Bifurcation diagrams showing saddles (black) and stable nodes (green). At least one saddle point exists whenever $\beta>1$.}
\label{bifstab}
\end{figure}

For $(a,b,k)=(2,8,2)$, we are in the shaded region of Figure~\ref{fig:alphabeta}. The dynamics are shown in Figure~\ref{bif1}, where the saddle has now replaced by  two saddles and a stable node, as a result of a saddle node (fold) bifurcation at $\beta = \beta_-$ (see Figure~\ref{bifstab}).

For $(a,b,k)=(1,8,2)$, we return to the unshaded region of Figure~\ref{fig:alphabeta}. We see the dynamics in Figure~\ref{bif2}, which is qualitatively similar to Figure~\ref{bif1}. Finally in Figure~\ref{bif3}, when $(a,b,k)=(1,1,2)$, we have $(\alpha,\beta) = (2,1)$, which corresponds to the case when the cubic equation \eqref{fixpoint} only has the trivial solution $X_e=0$.

What happens as we cross the curves $\beta = \beta_{\pm}$ in Figure~\ref{fig:alphabeta}? Figure \ref{bifstab}(a) shows the fold bifurcations that occur at $\beta_-$ and $\beta_+$ when $3<\alpha<4$. The black lines correspond to (unstable) saddles and the green line to a stable node. So if we fix $\alpha$ such that $3<\alpha<4$ and then vary $\beta$, there is no stable integrated population mix for $1<\beta<\beta_-$. Then for $\beta_-<\beta<\beta_+$, we can have a stable integrated population mix, although that all depends on the initial values of $X,Y$ (see Figure~\ref{basins}). Finally for $\beta>\beta_+$, we lose that stable solution via another fold bifurcation, and the remaining solution is unstable. 

In Figure \ref{bifstab}(b), when $\alpha>4$, we have a single fold at $\beta_-$. So if we now fix $\alpha$ in this range and vary $\beta$, there is no stable integrated population mix for $1<\beta<\beta_-$. But for $\beta>\beta_-$, we can have a stable integrated population mix, again depending on the initial values of $X,Y$ (see Figure~\ref{basins}).

It is only possible to have an integrated population mix when $\alpha  >  3$ from \eqref{alpha} (that is $ak  >  3$ from \eqref{ak}). In socio-economic terms, if $k=1$ for example, this result means that there must exist some population of each type that is content to live in up to a $3:1$ minority in order for a stable mixed state to be possible. So criterion \eqref{ak} is a generalisation to arbitrary $k$ of Schelling's observation \cite[p.172]{Schelling1971} that ``For straight-line tolerance schedules and equal numbers of the two colors, there is no intersection of the two parabolas unless the tolerance schedules have vertical intercepts of $3.0$, with median tolerance of $1.5$".

\section{Neighbourhood tipping and basins of attraction}\label{basins1area}
When discussing the time-evolution of the BNM, Schelling \cite[p.181]{Schelling1971} considers the possibility that "a recognizable new minority enters a neighbourhood in sufficient numbers to cause the earlier residents to begin evacuating". In dynamical systems terms, this means that a perturbation of the system may give rise to a different equilibrium point. In social science terms, this is the phenomenon known as {\it neighbourhood tipping}, which we now consider within the framework of dynamical systems.

The {\it basin of attraction} of a (necessarily stable) equilibrium $(X_e,Y_e)$ is the set of initial conditions $(X,Y)$ that lead to $(X_e,Y_e)$, under the action of the dynamical system. Unstable equilibria can not have basins of attraction (although the stable manifolds of saddles do divide phase space). Since our dynamical system is deterministic (no noise), each initial condition belongs to one basin of attraction. Thus neighbourhood tipping occurs when a perturbation moves the state of a system from one basin of attraction to another. 

\begin{figure}[ht]
\centering
\begin{subfigure}{.45\textwidth}
\centering
\includegraphics[width=\linewidth]{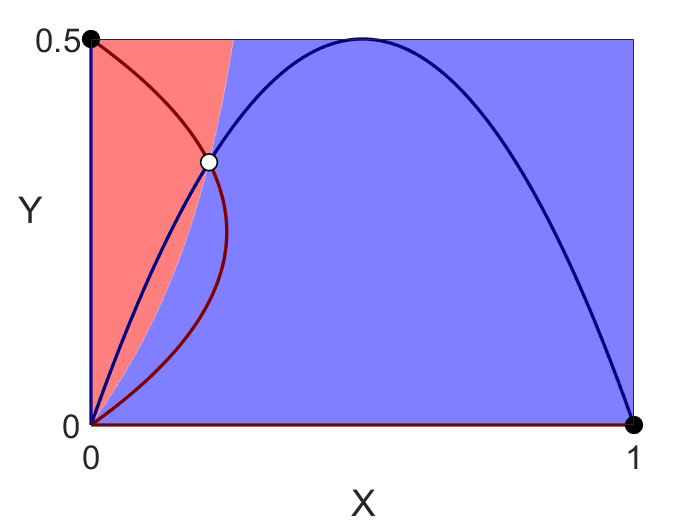}
\caption{$(a,b,k)=(2,2,2)$; $(\alpha,\beta) = (4,4)$}
\label{b1basins1}
\end{subfigure}\qquad\qquad
\begin{subfigure}{.45\textwidth}
\centering
\includegraphics[width=\linewidth]{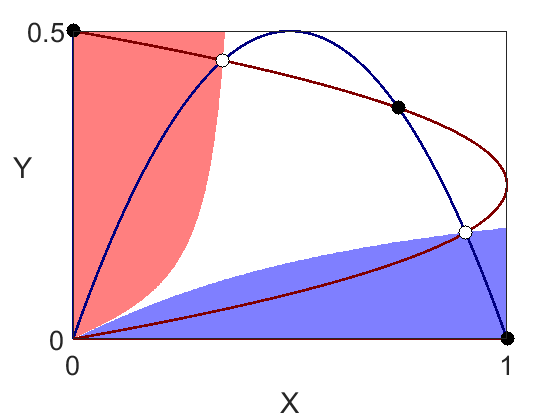}
\caption{$(a,b,k)=(2,8,2)$; $(\alpha,\beta) = (4,16)$}
\label{1basins2}
\end{subfigure} \\
\begin{subfigure}{.45\textwidth}
\centering
\includegraphics[width=\linewidth]{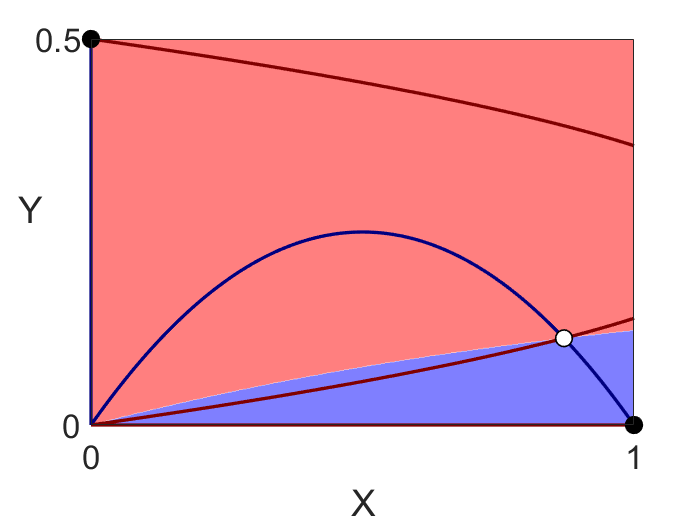}
\caption{$(a,b,k)=(1,8,2)$; $(\alpha,\beta) = (2,8)$}
\label{1basins3}
\end{subfigure}\qquad\qquad
\begin{subfigure}{.45\textwidth}
\centering
\includegraphics[width=\linewidth]{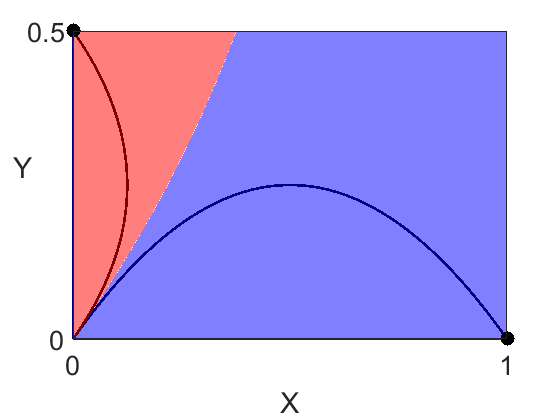}
\caption{$(a,b,k)=(1,1,2)$; $(\alpha,\beta) = (2,1)$}
\label{1basins4}
\end{subfigure}
\caption{Blue regions correspond to the basin of attraction of $X$-only equilibrium $(1,0)$, and red regions to the basin of attraction of the $Y$-only equilibrium $(0,\frac{1}{k})$. White denotes the basin of attraction of the stable mixed state (integrated population mix) obtained from \eqref{fixpoint} when it exists.}
\label{basins}
\end{figure}

In Schelling's work it is (tacitly) assumed that tipping points correspond to the boundaries of the parabolas (nullclines) of Figure~\ref{SchPP}. But this can not be the case owing to the presence of saddle points. We have computed the basins of attraction for the stable equilibria. Figure \ref{basins} shows these areas of phase space for the cases given in Figure~\ref{phaseports}.

\section{Limiting individual populations}\label{LET}
Schelling \cite[Figure 22, p173.]{Schelling1971} stated that ``limiting the numbers allowed to be present in the [neighbourhood] can sometimes produce a stable mixture". We now show exactly how such stable mixtures can be achieved.  
\begin{figure}[ht]
\centering
\includegraphics[width=0.4\linewidth]{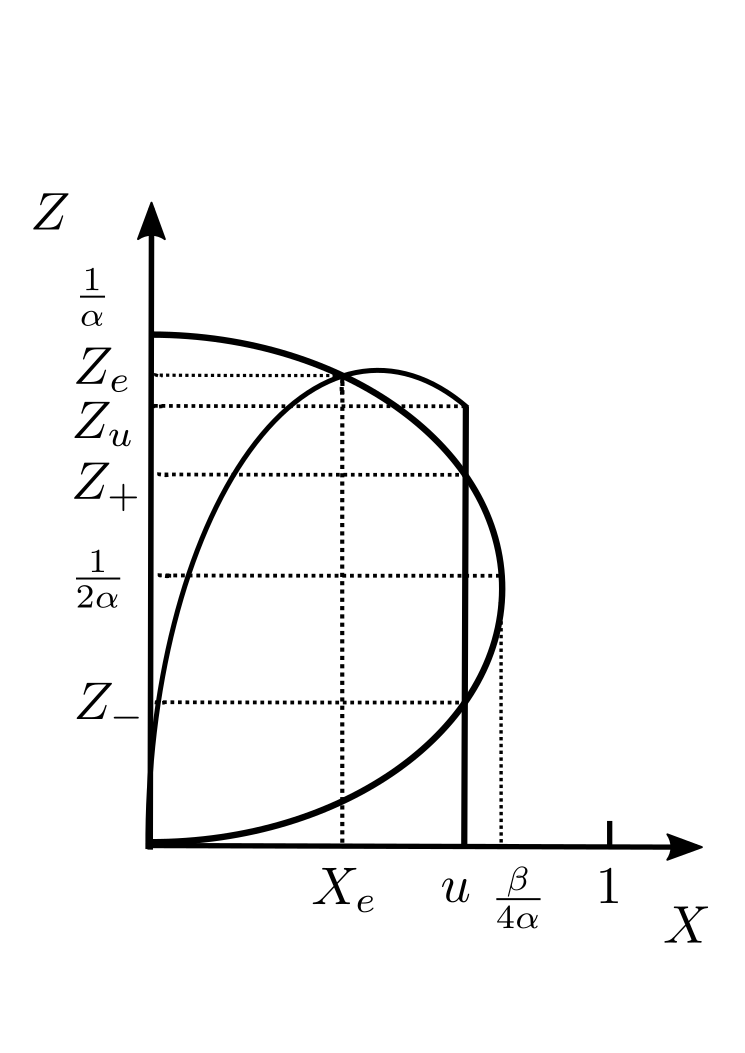}
\caption{Limiting the $X$-population: the case when $u< \frac{\beta}{4\alpha}$ and $Z_- < Z_+ < Z _u$.}
\label{fig:limitX}
\end{figure} 

\subsection{New equilibria}
Let us restrict the $X$-population to a maximum value of $u$, where $u \in (0,1)$, as shown in Figure~\ref{fig:limitX}. The $Z$-population is not restricted and so the corresponding maximum tolerance limit of the $X$-population is unchanged at $(X,Z)=(\frac{\beta}{4\alpha},\frac{1}{2\alpha})$. When the $X$-population is at its limiting value $X=u$, we have $Z=Z_u \equiv u(1-u)$. Then it is clear that we need to have $$u< \frac{\beta}{4\alpha}$$ for the limiting of the $X$-population to have any effect. Then the new population mixtures will correspond to the intersections $Z=Z_{\pm}$ of the line $X=u$ with the parabola $X=\beta Z(1-\alpha Z)$. Hence
\begin{equation}\label{Ypm}
Z_{\pm} = \frac{1}{2\alpha}\left [1 \pm \sqrt{1-\frac{4\alpha u}{\beta}} \right ].
\end{equation}
In general $Z_- \le Z_+$. If $Z_u < Z_- < Z _+$, no new intersections can be created by restricting the $X$-population and any existing equilibrium does not change. If $Z_- < Z_u < Z _+$, then no new intersections can be created by restricting the $X$-population and the existing equilibrium {\it will} change. But if $Z_- < Z_+ < Z _u$ (the case shown in Figure~\ref{fig:limitX}), then we may be able to produce new population mixes.

Our aim is to find a curve in the $(\alpha, \beta)$ plane which separates regions where a new stable integrated population mix is possible by limiting the $X$-population from regions where it is not. Points on this curve must satisfy
\begin{equation}\label{Xpopdivide}
X_e=u= \frac{\beta}{4\alpha}, \quad Z_e=Z_u=Z_-=Z_+ = \frac{1}{2\alpha},
\end{equation}
which happens when
\begin{equation}\label{Xpopbetacrit}
\beta = 2(\alpha \pm \sqrt{\alpha^2-2\alpha}).
\end{equation}

\begin{figure}[h!]
\centering
\begin{subfigure}{.45\textwidth}
\centering
\includegraphics[width=\linewidth]{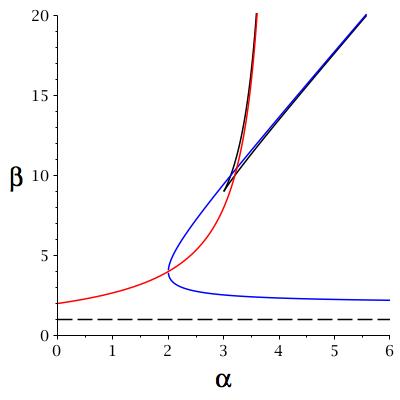}
\label{fig:alphabetalimitXY}
\end{subfigure}\qquad\qquad
\begin{subfigure}{.45\textwidth}
\centering
\includegraphics[width=\linewidth]{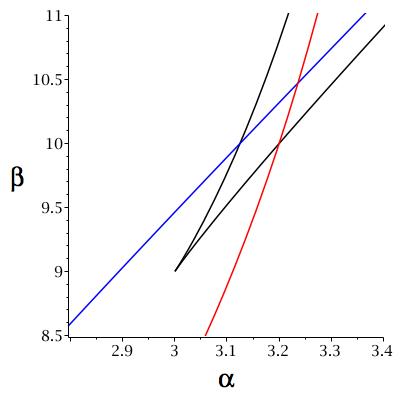}
\label{fig:alphabetalimitXYzoom}
\end{subfigure}
\caption{New population mixes can be created to the right of the blue curves, given by \eqref{Xpopbetacrit}, by restricting the $X$-population and to the left of the red curves, given by \eqref{Ypopbetacrit}, by restricting the $Z$-population. (a) The tolerance limits at the four points labelled (a), (b), (c) and (d) are shown in Figure~\ref{fig:limitXYexamples}. The black curves are from Figure~\ref{fig:alphabeta}. (b) Detail around the point $P$: $(\alpha,\beta)=(3,9)$.}
\label{fig:noref}
\end{figure}

Now suppose that we do not the restrict the $X$-population, but that the $Z$-population is restricted to $Z=v$, where $v \in (0,\frac{1}{\alpha})$. Similar reasoning as before shows that we need $$v<\frac{1}{4}$$ for the limiting of the $Z$-population to have any effect. When $Z=v$, we have that $X_v=\beta v(1-\alpha v)$ and the line $Z=v$ crosses the $X$-population tolerance limit where
\begin{equation}\label{Xpm}
X_{\pm} = \frac{1}{2}\left [1 \pm \sqrt{1-4v} \right ].
\end{equation}
Then the curve in the $(\alpha, \beta)$ plane which separates regions where a new stable integrated population mix is possible by limiting the $Z$-population from regions where it is not is given by
\begin{equation}\label{Ypopdivide}
X_e=X_u=X_-=X_+ = \frac{1}{\alpha}, \quad  Y_e=v= \frac{1}{4},
\end{equation}
which happens when
\begin{equation}\label{Ypopbetacrit}
\beta = \frac{8}{4-\alpha}.
\end{equation}
Results \eqref{Xpopbetacrit} and \eqref{Ypopbetacrit} are shown in Figure~\ref{fig:noref}. New population mixes can be created to the right of the blue curve, given by \eqref{Xpopbetacrit}, by restricting the $X$-population and to the left of the red curve, given by \eqref{Ypopbetacrit}, by restricting the $Z$-population.

\subsection{Stability of new equilibria}
Even though we may be able to create new population mixes by limiting a particular population, we do not know if that new mix is stable or not. We shall now answer that question. 

When $X=u$, the dynamics of the $Z$-population is governed by
\begin{eqnarray}\label{Ylimdyn}
\dot Z&=&\left[\beta Z(1-\alpha Z)-u\right]Z.
\end{eqnarray}
The stability of the equilibrium $Z=Z_{\pm}$ of \eqref{Ylimdyn} is determined by the eigenvalue
\begin{equation}
\lambda^u_{\pm} = \frac{\beta}{2\alpha}\left [-(1-\frac{4\alpha u}{\beta}) \mp \sqrt{1-\frac{4\alpha u}{\beta}} \right ].
\end{equation}
For $u \in [0,\frac{\beta}{4\alpha}]$, we can see that $\lambda^u_{\pm} \lessgtr 0$. So $Z_+$ is always stable and $Z_-$ is always unstable. Hence if we are able to restrict the $X$-population (so we choose $\alpha,\beta$ to the right of the blue curve in Figure~\ref{fig:noref}), we will always produce a new population mix that is always stable.

When $Z=v$, the dynamics of the $X$-population is governed by
\begin{eqnarray}\label{Xlimdyn}
\dot X&=&\left[X(1-X)-v\right]X.
\end{eqnarray}
The stability of the equilibrium $X=X_{\pm}$ of \eqref{Xlimdyn} is determined by the eigenvalue
\begin{equation}
\lambda^v_{\pm} = \frac{1}{2}\left [-(1-4v) \mp \sqrt{1-4v} \right ].
\end{equation}
For $v \in [0,\frac{1}{4}]$, we can see that $\lambda^v_{\pm} \lessgtr 0$. So $X_+$ is always stable and $X_-$ is always unstable. Hence if we are able to restrict the $Z$-population (so we choose $\alpha,\beta$ to the left of the red curve in Figure~\ref{fig:noref}), we will always produce a new population mix that is always stable.

\subsection{Tolerance limits}
\begin{figure}[h!]
\centering
\begin{subfigure}{.45\textwidth}
\centering
\includegraphics[width=\linewidth]{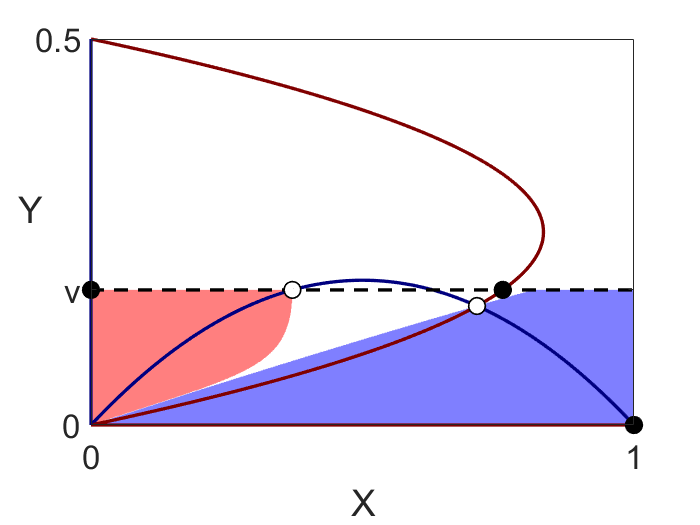}
\caption{$(\alpha,\beta) = (1.5,5)$; $(a,b,k)=(\frac{3}{4},\frac{20}{3},2)$.}
\label{fig:alpha1p5beta5limitXY}
\end{subfigure}\qquad\qquad
\begin{subfigure}{.45\textwidth}
\centering
\includegraphics[width=\linewidth]{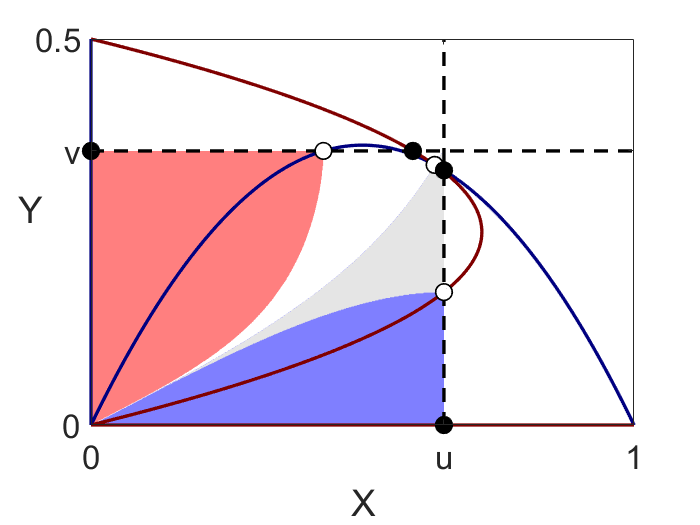}
\caption{$(\alpha,\beta) = (2.9,8.35)$; $(a,b,k)=(1.45,5.76,2)$.}
\label{fig:alpha2p9beta8p35limitXY}
\end{subfigure} \\
\begin{subfigure}{.45\textwidth}
\centering
\includegraphics[width=\linewidth]{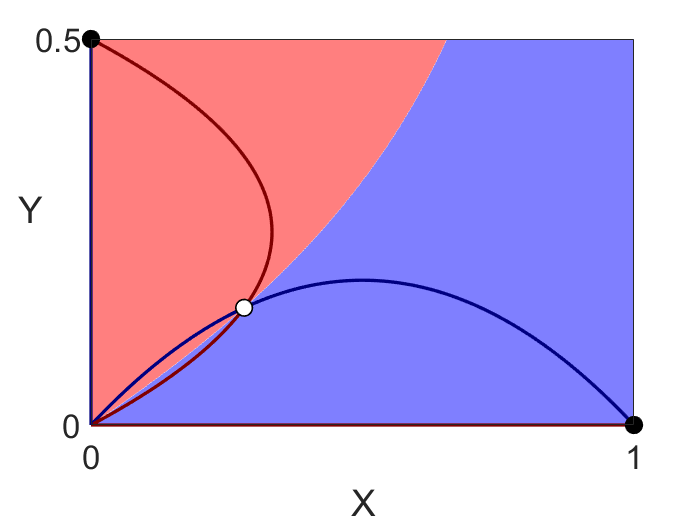}
\caption{$(\alpha,\beta) = (1.5,2)$; $(a,b,k)=(\frac{3}{4},\frac{8}{3},2)$.}
\label{fig:alpha1p5beta2limitXY}
\end{subfigure}\qquad\qquad
\begin{subfigure}{.45\textwidth}
\centering
\includegraphics[width=\linewidth]{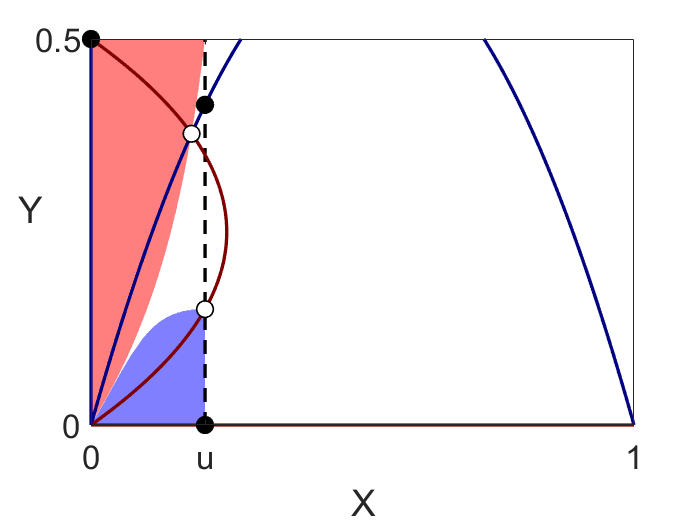}
\caption{$(\alpha,\beta) = (5,5)$; $(a,b,k)=(\frac{5}{2},2,2)$.}
\label{fig:alpha5beta5limitXY}
\end{subfigure}
\caption{The tolerance limits at each of the four points (a), (b), (c) and (d) of Figure~\ref{fig:noref}(a), 
with candidate cut-off values shown as a black dashed lines. 
Note that $k=2$ in all four examples, and basins of attraction are coloured as in Figure~\ref{basins}, with gray used to denote the basin of attraction of an additional stable mixed state.}
\label{fig:limitXYexamples}
\end{figure}

The tolerance limits are shown in Figure~\ref{fig:limitXYexamples}, at each of the four points (a), (b), (c) and (d) of figure~\ref{fig:noref}(a). In each case, when both populations are unrestricted, the only stable equilibria are fully segregated; there are no stable integrated population mix. In this section, we demonstrate how a limit on population can produce a new stable population mix.

In Figure~\ref{fig:limitXYexamples}, 
we have introduced candidate cut-off values $X=u$ and $Y=v$ are shown as black dashed lines. In order to introduce new fixed points, we must have $u<\frac{1}{2\alpha}R_Y(\frac{1}{2\alpha})$ or $v<\frac 12 R_X(\frac 12)$, i.e. the limit of one type must intersect the nullcline of the other type. There are two such lines in figure~\ref{fig:alpha2p9beta8p35limitXY} and one each in figures~\ref{fig:alpha1p5beta5limitXY} and~\ref{fig:alpha5beta5limitXY}. Basins of attraction are coloured as in figure~\ref{basins} and, in figure~\ref{fig:alpha1p5beta2limitXY}, we have $2$ stable mixed states, with basins of attraction coloured white and gray.

In fact, we can in fact create up to $7$ new equilibria as a result of limiting numbers when $\alpha$ and $\beta$ lie in the lozenge-shaped region containing in figure~\ref{fig:noref}(b), by choosing $u$ and $v$ accordingly. 

\section{Different tolerance schedules}\label{VTS}
The use of linear tolerance schedules, as in Figure~\ref{fig:tol} simplifies the analysis. But quantitative results \cite{Clark1991} show that modifications to the linear tolerance schedule may better describe the social context. 
Schelling \cite[Figures 25-29]{Schelling1971} also used nonlinear tolerance schedules.

In this section, we show how our approach can be modified to take account of different tolerance schedules and study their dynamics. All possible tolerance schedules must satisfy $R_X(0)=1,\ R_X(1)=0,\ R_Y(0)=\beta,\ R_Y(\frac{1}{k})=0$, in terms of the original populations $X,Y$.

After suitable scalings, our Schelling dynamical system for general tolerance schedules $R_X(X), R_Z(Z)$ can be written as
\begin{eqnarray}
\dot X&=&\left[XR_X(X)-Z\right]X\\
\dot Z&=&\left[ZR_Z(Z)-X\right]Z. \nonumber
\end{eqnarray}
To find equilibria solutions $(X,Z)=(X_e,Z_e)$, we must find positive solutions of
\begin{eqnarray}
Z_e&=&X_eR_X(X_e),\\
X_e&=&Z_eR_Z(Z_e). \nonumber
\end{eqnarray}
To examine the stability of these equilibria, we must calculate the eigenvalues of the Jacobian
\begin{eqnarray}
J(X,Z)=\left( \begin{array}{cc}
2XR_X(X)+X^2\frac{dR_X(X)}{dX}-Z & -X \\
-Z & 2ZR_Z(Z)+Z^2\frac{dR_Z(Z)}{dZ}-X \end{array} \right),
\end{eqnarray}
evaluated at $(X,Z)=(X_e,Z_e)$. Stable population mixes will have both eigenvalues negative.

\subsection{Polynomial functions}
Departing from linear tolerance schedules, our simplest choice is a polynomial function. Consider the following:
\begin{align}
R_X(X)=&RP_X^p(X)\equiv\left(1-X^p\right),\\
R_Z(Z)=&RP_Z^q(Z)\equiv\beta\left(1-\alpha^qZ^q\right), \nonumber
\end{align}
where $p,q \in \mathbb{N}^+$ are positive integers. For $p=q=1$ we have the linear tolerance schedules \eqref{nonDS}. Larger values of $p,q$ give more tolerant populations.

To find equilibria, we must solve the following equation for $X_e \ge 0$:
\begin{equation}\label{fixpointpoly}
(-1)^{q+1}X_e^{p(q+1)+q}+\sum_{k=1}^q (-1)^k \binom {q+1} {k} X_e^{pk+q} \\
+X_e^q+\frac{1}{\alpha^q}X_e^p+\frac{1-\beta}{\alpha^q\beta}=0
\end{equation}
and then ensure that $Z_e = X_e(1-X_e^p)$ is also positive. When $p=q=1$, \eqref{fixpointpoly} reduces to \eqref{fixpoint}.

Since \eqref{fixpointpoly} has $p(q+1)+q$ roots, it would appear that this choice of tolerance schedule could give us more non-trivial equilibria for increased $p,q$. However Descartes' rule of signs applied to \eqref{fixpointpoly} shows us that there are {\it at most} $q+2$ real positive values of $X_e$. But for several of these solutions, we find $Z_e < 0$. In fact, since the tolerance schedules are generalised parabolae, it turns out that we have only a maximum of three solutions of \eqref{fixpointpoly} where both $X_e>0$ and $Z_e>0$. Numerical methods must be used to find these solutions.

Another choice of polynomial is possible, which results in a \textit{less tolerant population}. Consider
\begin{align}
R_X(X)=&RQ_X^r(X)\equiv\left(1-X\right)^r\\
R_Z(Z)=&RQ_Z^s(Z)\equiv\beta\left(1-\alpha Z\right)^s. \nonumber
\end{align}
As before, the equilibria are those roots of a high order polynomial that satisfy both $X_e>0$ and $Z_e>0$. This polynomial is found using the multinomial theorem and its roots must be calculated numerically. As before, we observe that we only have a maximum of three such solutions. Hence the change to a polynomial tolerance schedule does not produce extra integrated equilibria.

Note that the choice of functions $RP_X^p(X), RP_Z^q(Z), RQ_X^r(X), RQ_Z^s(Z)$ are not unique.

\subsection{Exponential functions}
Another possible tolerance schedule involves the use of exponential functions. Consider
\begin{align}
R_X(X)=&RE_X^{\mu}(X)\equiv\frac{1}{1-e^{-\mu}}\left(e^{-\mu X}-e^{-\mu}\right)\label{expTS}\\
R_Z(Z)=&RE_Z^{\nu}(Z)\equiv\frac{\beta}{1-e^{-\frac{\nu}{\alpha}}}\left(e^{-\nu Z}-e^{-\frac{\nu}{\alpha}}\right),\nonumber
\end{align}
where $\mu,\nu>0$. Equilibria and their stability properties have to be calculated numerically.

\begin{figure}[ht]
	\centering
	\begin{subfigure}{.45\textwidth}
		\centering
		\includegraphics[width=\linewidth]{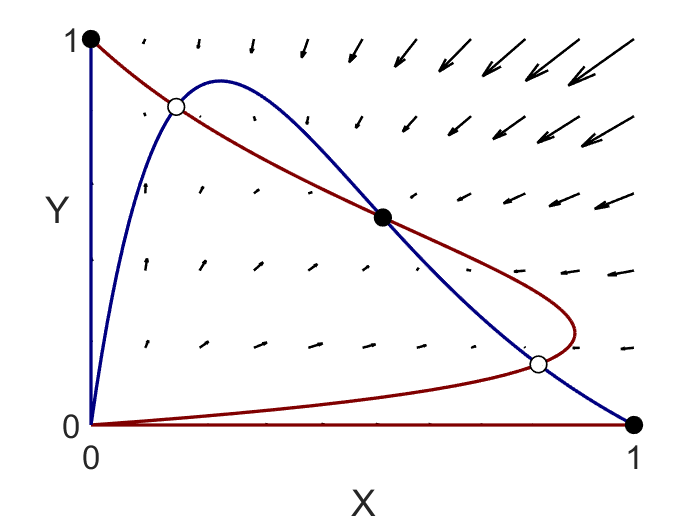}
		\caption{Phase portrait with $RE_X^4,\ RE_Y^4,\ a=b=10,\ k=1$}
		\label{Ex1}
	\end{subfigure}
	\begin{subfigure}{.45\textwidth}
		\centering
		\includegraphics[width=\linewidth]{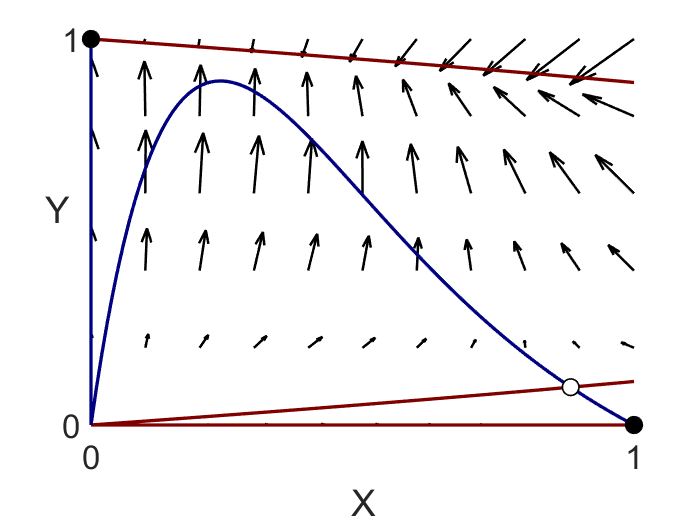}
		\caption{Phase portraits with $RE_X^4,\ RP_Y^1,\ a=10,\ b=10,\ k=1$}
		\label{Ex2}
	\end{subfigure}
	\caption{Phase portraits demonstrating that a globally more tolerant minority population can eliminate a stable mixed state}
	\label{PPexamples}
\end{figure}

Figure \ref{PPexamples} illustrates the effect of exponential tolerance schedules, in terms of the original $X,Y$ populations. In Figure \ref{Ex1}, we take $RE_X^4$ and $RE_Y^4$, with $a=b=10$ and $k=1$. In Figure \ref{Ex2} we revert to a linear tolerance schedule for the $Y$-population, taking $RE_X^4,\ RP_Y^1,\ a=b=10,\ k=1$. We observe that changing the tolerance schedule of {\it one} population, so that they become more tolerant, can result in the loss of a stable mixed state. We can understand this phenomenon by considering an initial condition $(X_0,Y_0)$ that lies close to the peak of the $X$-nullcline. The key is that $\dot Y$ has changed sign from negative to positive at $(X_0,Y_0)$. We can interpret this intuitively as follows: "some $Y$s wanted to leave, but now they're more tolerant, they're happy to stay. In fact, $Y$s are so happy that $Y$s come in, which makes $X$s want to leave".

For further examples of phase portraits with different tolerance schedules, the reader is referred to the PhD thesis of the first author
, available at \href{https://www.imperial.ac.uk/people/d.haw/research.html}{D. Haw's PhD thesis}.

\section{Discussion, conclusions and further work}\label{sec:disc}
We have taken Schelling's \cite{Schelling1969, Schelling1971} bounded neighbourhood model (BNM) and derived a (scaled) dynamical system \eqref{nonDS} that allows for phase-plane analysis. In doing so, we can identify terms in the governing equations and attach social meaning to each of them. The details vary slightly between different tolerance schedules, but the basic principles can be illustrated by using the linear tolerance schedule. In this case, the dynamical system \eqref{nonDS} is reproduced here in an expanded form for convenience.
\begin{eqnarray}
\dot X&=& X^2-X^3-XZ \label{expnonDS}\\
\dot Z&=& \beta Z^2-\alpha \beta Z^3-XZ. \nonumber
\end{eqnarray}

Let us consider the first equation in \eqref{expnonDS}. The term $X^2$ represents growth of the $X$-population, as it enters the neighbourhood, unhindered by lack of space or the presence of the $Z$-population. It is a measure of the intrinsic popularity of the neighbourhood. The term $X^3$ represents a decay in the $X$-population, corresponding to a reduction in available space in the neighbourhood as the population expands, brought about by the finite size of the neighbourhood. The term $XZ$ represents a decay in the $X$-population induced by the presence of $Z$-population. Similar considerations apply to the $Z$-population dynamics, given by the second equation in \eqref{expnonDS}.

In summary, the dynamics of one population in the presence of another can be summarised as follows
\begin{multline}
\{\mbox{rate of population change}\} = \{\mbox{intrinsic popularity of neighbourhood}\} - \{\mbox{finite size of neighbourhood}\} \\
-\{\mbox{presence of other population}\}
\end{multline}
The precise form of these terms could be determined by field experiments. By approaching the dynamics from this perspective, we have a complementary approach to that of the tolerance schedule.

The categorisation of all possible equilibria and the partitioning of phase-space into basins of attraction allow us describe the dynamics of segregation with respect to key parameters of the system. In particular, saddle points are a common feature of dynamics, and yield dynamics that are fundamentally different to the intuitive description given in Schelling's paper.

Many additional variants of the BNM are possible, including the use of other non-linear tolerance schedules, limiting the total population present, and limiting the ration $X/Y$. The analytic intractability of such systems means that a computational approach is necessary.

Further work involves generalising Schelling's BNM to describe systems of neighbouring geographical areas and the flow of populations between them. Also, access to demographic data with multiple time-points may help to derive realistic parameter estimates and tolerance schedules.

\section{Acknowlegements}

We would like to thank Professor Simon Burgess of the University of Bristol for first suggesting the development of Schelling's Bounded Neighbourhood Model.


\newpage
\bibliographystyle{plain}

\begin{thebibliography}{10}

\bibitem{Dodson2014}
{A. Dodson}.
\newblock {Using the CoSMoS Approach to study Schelling's Bounded Neighbourhood
  Model}.
\newblock In {\em CoSMoS}, pages 1--12, 2014.

\bibitem{Benenson1998}
Itzhak Benenson.
\newblock {Multi-agent simulations of residential dynamics in the city}.
\newblock {\em Computers, Environment and Urban Systems}, 22(1):25--42, 1998.

\bibitem{Benenson1999}
Itzhak Benenson.
\newblock {Modeling population dynamics in the city: From a regional to a
  multi-agent approach}.
\newblock {\em Discrete Dynamics in Nature and Society}, 3(2-3):149--170, 1999.

\bibitem{Benenson2013}
Itzhak Benenson.
\newblock {\em {Agent-based models of geographical systems}}, volume~27.
\newblock 2013.

\bibitem{Benenson2002}
Itzhak Benenson, Itzhak Omer, and Erez Hatna.
\newblock {Entity-based modeling of urban residential dynamics: The case of
  Yaffo, Tel Aviv}.
\newblock {\em Environment and Planning B: Planning and Design},
  29(4):491--512, 2002.

\bibitem{Clark1991}
W.~A.V. Clark.
\newblock {Residential preferences and neighborhood racial segregation: A test
  of the Schelling segregation model}.
\newblock {\em Demography}, 28:1--19, 1991.

\bibitem{Fossett2006a}
Mark Fossett.
\newblock {\em {Ethnic Preferences, Social Distance Dynamics, and Residential
  Segregation: Theoretical Explorations Using Simulation Analysis}}, volume~30.
\newblock 2006.

\bibitem{Fossett2005}
Mark Fossett and Warren Waren.
\newblock {Overlooked implications of ethnic preferences for residential
  segregation in agent-based models}.
\newblock {\em Urban Studies}, 42(11):1893--1917, 2005.

\bibitem{Grauwin2012}
S~Grauwin, F~Goffette-Nagot, and P~Jensen.
\newblock {Dynamic models of residential segregation: An analytical solution}.
\newblock {\em Journal of Public Economics}, 96:124--141, 2012.

\bibitem{Henry2011a}
Adam~Douglas Henry, Pawe{\l} Pra{\l}at, and Cun-Quan Zhang.
\newblock {Emergence of Segregation in Evolving Social Networks.}
\newblock {\em Proceedings of the National Academy of Sciences of the United
  States of America}, 108(21):8605--8610, 2011.

\bibitem{Laurie2003}
Alexander Laurie and Narendra Jaggi.
\newblock {Role of `vision' in neighbourhood racial segregation: A variant of
  the Schelling Segregation Model}.
\newblock {\em Urban Studies}, 40(13):2687--2704, 2003.

\bibitem{Pancs2007}
Romans Pancs and Nicolaas J N~J Vriend.
\newblock {Schelling's spatial proximity model of segregation revisited}.
\newblock {\em Journal of Public Economics}, 91(487):1--24, feb 2007.

\bibitem{Pollicott2001}
Mark Pollicott and Howard Weiss.
\newblock {The Dynamics of Schelling-Type Segregation Models and a Nonlinear
  Graph Laplacian Variational Problem}.
\newblock {\em Advances in Applied Mathematics}, 27(1):17--40, jul 2001.

\bibitem{Santos2017}
Miguel Santos and Jorge Pe.
\newblock {Antisocial rewarding in structured populations}.
\newblock {\em bioRxiv}, pages 1--29, 2017.

\bibitem{Schelling1969}
T.~C. Schelling.
\newblock {Models of segregation}.
\newblock {\em The American Economic Review}, 59:488--493, 1969.

\bibitem{Schelling1971}
T.~C. Schelling.
\newblock {Dynamic models of segregation}.
\newblock {\em Journal of Mathematical Sociology}, 1:143--186, 1971.

\bibitem{Singh2009}
Abhinav Singh, Dmitri Vainchtein, and Howard Weiss.
\newblock {Schelling's segregation model: Parameters, scaling, and
  aggregation}.
\newblock {\em Demographic Research}, 21:341--366, 2009.

\bibitem{Singh2011a}
Abhinav Singh, Dmitri Vainchtein, and Howard Weiss.
\newblock {Limit sets for natural extensions of Schelling's segregation model}.
\newblock {\em Communications in Nonlinear Science and Numerical Simulation},
  16(7):2822--2831, 2011.

\bibitem{Hatna2012}
Artificial Societies and Social Simulation.
\newblock {Erez Hatna and Itzhak Benenson ( 2012 ) The Schelling Model of
  Ethnic Residential Dynamics : Beyond the Integrated - Segregated Dichotomy of
  Patterns}.
\newblock 15(2012):1--23, 2012.

\bibitem{StrogatzBook}
S.~H. Strogatz.
\newblock {\em {Nonlinear Dynamics and Chaos: With Applications to Physics,
  Biology, Chemistry and Engineering}}.
\newblock Addison--Wesley, January 1994.

\bibitem{Yizhaq2004}
Hezi Yizhaq, Boris~A. Portnov, and Ehud Meron.
\newblock {A mathematical model of segregation patterns in residential
  neighbourhoods}.
\newblock {\em Environment and Planning A}, 36(1):149--172, 2004.

\bibitem{Zhang2009}
Junfu Zhang.
\newblock {Tipping and Residential Segregation: A United Schelling Model}.
\newblock {\em Journal of Regional Science}, 51(1):167--193, 2009.

\end{thebibliography}

\end{document}